\documentclass[aps,twocolumn,showpacs,superscriptaddress,amsmath,amssymb]{revtex4}

\usepackage{graphicx}  
\usepackage{dcolumn}   
\usepackage{bm}

\begin{document}

\title{Origin of the excitonic recombinations in hexagonal boron nitride by spatially resolved cathodoluminescence spectroscopy}

\author{P. Jaffrennou}
\affiliation{Laboratoire d'Etude des Microstructures, 
ONERA-CNRS, BP 72, 92322 
Ch\^atillon Cedex, France}
\affiliation{D\'epartement de Mesures Physiques, 
ONERA, Chemin de la Huni\`ere, 91761 Palaiseau Cedex, France}
\affiliation{Laboratoire de Photonique Quantique et Mol\'eculaire, 
Institut d'Alembert, Ecole Normale Sup\'erieure de Cachan,
61 avenue du Pr\'esident Wilson 94235 Cachan Cedex, France}
\author{J. Barjon}
\affiliation{Groupe d'Etude de la Mati\`ere Condens\'ee
CNRS-Universit\'e de Versailles Saint-Quentin-en-Yvelines, 
1 place Aristide Briand, 92195 Meudon Cedex, France}
\author{J.-S. Lauret}
\affiliation{Laboratoire de Photonique Quantique et Mol\'eculaire, 
Institut d'Alembert, Ecole Normale Sup\'erieure de Cachan
61 avenue du Pr\'esident Wilson 94235 Cachan Cedex, France}
\author{B. Attal-Tr\'etout}
\affiliation{D\'epartement de Mesures Physiques, 
ONERA, Chemin de la Huni\`ere, 91761 Palaiseau Cedex, France}
\author{F. Ducastelle}
\email{francois.ducastelle@onera.fr}
\author{A. Loiseau}
\affiliation{Laboratoire d'Etude des Microstructures, 
ONERA-CNRS, BP 72, 92322 
Ch\^atillon Cedex, France}

\date{\today}

\begin{abstract}
The excitonic recombinations in hexagonal boron nitride (hBN) are investigated with spatially resolved cathodoluminescence spectroscopy in the UV range. Cathodoluminescence images of an individual hBN crystallite reveals that the 215 nm free excitonic line is quite homogeneously emitted along the crystallite whereas the 220 nm and 227 nm excitonic emissions are located in specific regions of the crystallite. Transmission electron microscopy images show that these regions contain a high density of crystalline defects. This suggests that both the 220 nm and 227 nm emissions are produced by the recombination of excitons bound to structural defects.
\end{abstract}


\maketitle

Hexagonal Boron Nitride (hBN) is a wide band gap semiconductor \cite{Blase95, Rubio94, Arnaud06, Wirtz05} isostructural with graphite. For a few years, much interest has been devoted to BN materials since hBN and more specifically BN nanotubes \cite{Wu04, Zhi05, Arenal05, Lauret05, Chen06, Berzina05, Berzina06, Jaffrennou07, Jaffrennoubis07} are expected to be promising materials for optoelectronic applications. Recently, experimental and theoretical studies have been undertaken in order to investigate the optical properties of hBN. Using luminescence experiments, the near band-edge emission has been observed in the UV region \cite{Taylor94, Watanabe04, Watanabe06, Watanabebis06, Silly07}. Theoretical groups have interpreted this UV emission as due to unusually strong excitonic effects \cite{Arnaud06, Wirtz05}. They propose that excitons in hBN are closer to Frenkel-type excitons with large binding energy (0.7 eV) than to the usual Wannier-like excitons of classical III-N semiconductors.

The hBN near-band edge emission consists in several peaks at 215 nm (5.77 eV), 220 nm (5.63 eV) and 227 nm (5.46 eV) \cite{Watanabe04}. Watanabe \textit{et al.} \cite{Watanabe06} recently observed that this UV band shows drastic changes after deformation of a single crystal \cite{Watanabe06, Watanabebis06}. By pressing a hBN single crystal between two fingers, the authors show that the relative intensities of the 215 nm  and 227 nm bands are reversed. After deformation, the CL intensity of the 215 nm band becomes negligible as compared to the 227 nm band which is then predominant. According to recent calculations \cite{Arnaud06}, the 215 nm luminescence band has been attributed to Frenkel-type free excitonic recombinations and Watanabe \textit{et al.} tentatively assign the 227 nm band, which appears predominantly after deformation, to excitons bound to stacking faults or to the shearing of lattice planes \cite{Watanabe06}. 

In this work, we present a detailed investigation of the excitonic luminescence in hBN. By combining spatially resolved cathodoluminescence (SR-CL) spectroscopy with a structural analysis of the crystal by means of Transmission Electron Microscopy (TEM) and Selected Area Electron Diffraction (SAED) analyses, we investigate the nature of the 220 nm and 227 nm excitonic recombinations in terms of excitons bound to well identified structural defects. 

In these experiments, commercial Aldrich hBN powders are used. The samples were dispersed in ethanol and deposited on carbon coated TEM copper grids in order to isolate individual hBN crystallites. Commercial Starck hBN powders were also investigated and gave comparable results (not reported here). TEM and SAED analyses were performed with a Philips CM20 transmission electron microscope at an accelerating voltage of 200 keV. From the transparency of the crystallite to electrons, one can guess that the thickness of the sample is under 200 nm. For the CL experiments, samples were imaged at 102 K using a 20 keV, 10 nA electron beam in a JEOL840 scanning electron microscope. The spatial resolution of the CL images is then limited by the beam size of about 0.4 $\mu$m. A Horiba Jobin Yvon SAS system is used to analyze the CL emission in the UV range. The light is collected with a parabolic mirror and reflected using metallic optics into a TRIAX550 monochromator equipped with a UV-enhanced silicon CCD camera to get the CL spectra, or a Hamamatsu (R943-02) photomultiplier to collect the CL images.  

\begin{figure}[ht]
	\centering
		\includegraphics[width=8cm]{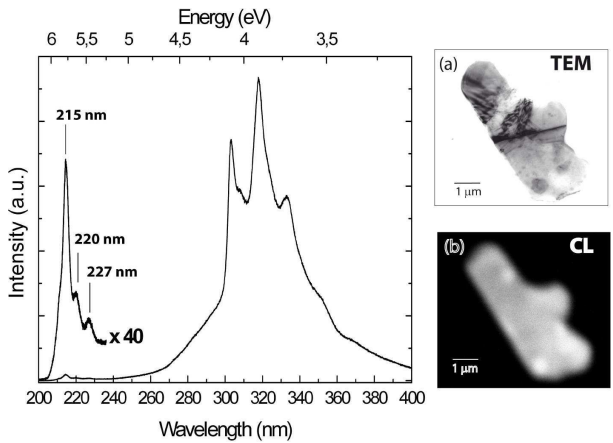}
	\caption{CL spectrum of the hBN crystallite, T=100 K, I=10 nA. (a) TEM image of the hBN crystallite and (b) Polychromatic CL image of the hBN crystallite }
	\label{fig:spectreCL}
\end{figure}

\begin{figure}[ht]
	\centering
		\includegraphics[width=8cm]{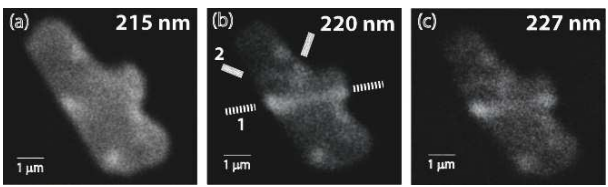}
	\caption{Monochromatic CL images of the hBN crystallite at (a) 215 nm, (b) 220 nm and (c) 227 nm ;  T=100 K}
	\label{fig:imagesCLhBN}
\end{figure}

The individual hBN crystal studied here is about 5 $\mu$m large and is shown in Figure \ref{fig:spectreCL}a. This image is a bright field TEM image of the crystallite, oriented in the (001) zone axis on the thin central area (dark contrast). TEM analysis of this object indicates that the crystallite is flat, with thicker zones corresponding to the superimposition of other small crystallites under the main one (thickness contrast in TEM). The polychromatic (i.\ e. total cathodoluminescence intensity) CL image (Figure \ref{fig:spectreCL}b) of the crystallite shows that it is strongly luminescent and that the total intensity is rather homogeneous in the whole crystal, with slightly stronger luminescence spots observed in the thicker zones of the crystal.  Its CL spectrum exhibits two bands : a near-band edge (maximum at 215 nm) and an impurity or defect-related luminescence band (maximum at 317 nm) with relative intensities comparable to previous observations \cite{Silly07, Museur07, Watanabe07}. The deep blue impurity or defect emission band was recently assigned to intrinsic impurity centers such as C and O impurities \cite{Watanabe07}. The excitonic recombinations are composed of three peaks at 215 nm, 220 nm and 227 nm. Monochromatic (i.\ e. filtered at a specific wavelength) CL images of the same object have been recorded at the three wavelengths corresponding to the lines in the excitonic luminescence band (Figure \ref{fig:imagesCLhBN}). 

At 215 nm (Figure \ref{fig:imagesCLhBN}a), the spatial distribution of the light emission over the crystallite is comparable to the polychromatic CL image (Figure \ref{fig:spectreCL}b) and exhibits a homogeneous spatial distribution. On the contrary, the monochromatic CL images filtered at 220 nm and 227 nm do not show such a distribution (Figure \ref{fig:imagesCLhBN}b and  \ref{fig:imagesCLhBN}c). In both cases, the emission is mainly localized on a bright line crossing the middle of the crystallite (Figure \ref{fig:imagesCLhBN}b : label 1), and dividing it into two parts. One could also distinguish two other weaker bright lines (Figure \ref{fig:imagesCLhBN}b : label 2) also emitting at these wavelengths. 

\begin{figure}[ht]
	\centering
		\includegraphics[width=8cm]{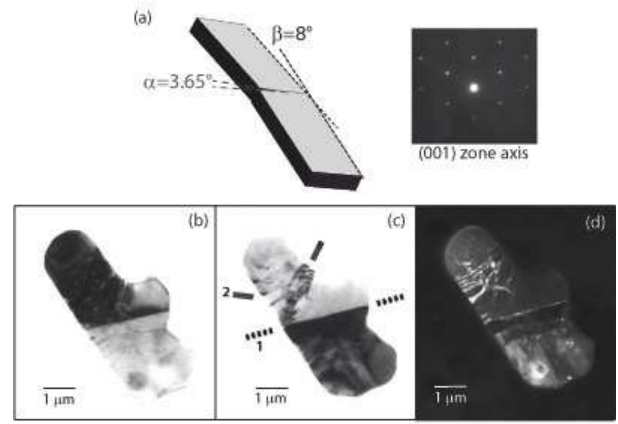}
	\caption{(a) Sketch of the hBN crystallite with Bright Field TEM images oriented in the (001) zone axis (b and c). (d) Dark Field TEM image. Inset : SAED pattern in the(001) zone axis}
	\label{fig:hBNBrightField}
\end{figure}

In order to identify the crystalline structure of the emitting lines observed in monochromatic CL images filtered at 220 nm and 227 nm, TEM analysis of the same object has been undertaken. First, by orienting the crystallite in the (001) zone axis, we have determined that it is made of two single crystalline domains separated by an interface depicted by the label 1 in Figure \ref{fig:hBNBrightField}. The $\alpha$ and $\beta$ tilt angles characterizing the misorientation between the two grains are respectively equal to 3.65$^\circ$ and 8$^\circ$, as schematized in Figure \ref{fig:hBNBrightField}a. Bright Field TEM images in Figure \ref{fig:hBNBrightField}b and \ref{fig:hBNBrightField}c correspond to orientations where one of the two grains is in exact (001) orientation and is imaged by a strong double contrast. Furthermore, one can distinguish several defects within each grain such as dislocations which are assembled into arrays and aligned along definite crystalline directions (label 2 in Figure \ref{fig:hBNBrightField}). Such an array is well visible in the weak beam image of Figure \ref{fig:hBNBrightField}d obtained with a reflection (150) of the (001) zone axis. By comparing these images to the monochromatic CL images at 220 nm and 227 nm (Figure \ref{fig:imagesCLhBN}b and \ref{fig:imagesCLhBN}c), it is remarkable that the luminescence is exactly localized at the grain boundary (Figure \ref{fig:hBNBrightField}c : label 1) and at the dislocations (Figure \ref{fig:hBNBrightField}c : label 2).

This analysis, based on the correlation between monochromatic CL images and TEM images, shows that the emissions at 220 nm and 227 nm are related to these structural defects: grain boundaries and dislocations. Such a spatial localization of the emitted light has already been observed in the luminescence from excitons bound to defects or impurities in III-V or II-VI semiconductors \cite{Mitsui96, Sun02, Bertram04}. In our case, the CL images filtered at 220 nm and 227 nm display the same spatial light distribution. The excitonic transitions responsible for these emissions can therefore be of the same type and we assign them to excitons bound to the same type of structural defects. These luminescence lines can then be interpreted either as two excitonic levels of the same exciton bound to structural defects or as phonon replica, the 227 nm lines being the first phonon replica of the bound excitonic transition at 220 nm. 

These results are consistent with previous works of Watanabe \textit{et al.} \cite{Watanabe06} on deformed hBN single crystal. It is noticeable that grain boundaries and dislocations were introduced by a mechanical stress in the case of the Watanabe \textit{et al.} study \cite{Watanabe06} whereas in our case, the defects are certainly due to thermal stress introduced during the crystal growth. 

To summarize, in this Letter, we have observed the hBN luminescence of free excitons emitting at 215 nm and of lower energy emissions at 220 nm and 227 nm which are localized on the hBN crystallite. Monochromatic CL images and TEM images on the same hBN crystallite have been compared in order to correlate the localization of the excitonic luminescence with the crystalline structure of the crystallite and the structural defects of hBN. The 220 nm and 227 nm luminescence bands could be assigned to excitons bound to grain boundaries and dislocations. Further investigation of hBN luminescence by means of photoluminescence spectroscopy and time resolved spectroscopy is under progress.

The authors are grateful to Thomas Schmid for technical support and to Andrei Kanaev and Luc Museur for helpful discussions and for communication of their unpublished results. LEM is a ``Unit\'e
mixte" ONERA-CNRS (UMR104). LPQM is a ``Unit\'e mixte" de
recherche associ\'ee au CNRS (UMR8537). GEMAC is a ``Unit\'e mixte"
de recherche associ\'ee au CNRS (UMR8635). This work has been
supported by the GDR-E ``nanotube" (GDRE2756).

\end{document}